\documentclass[a4paper,11pt]{article}
\usepackage{pos}
\usepackage{wrapfig}
\usepackage{adjustbox}
\usepackage{physics}
\usepackage{lineno}

\begin{document}

\title{Realtime Follow-up of External Alerts with the IceCube Supernova Data Acquisition System
}
 \ShortTitle{External alerts with IceCube SNDAQ system}

\author{Nora Valtonen-Mattila for the IceCube Collaboration} 





\emailAdd{nora.valtonen-mattila@icecube.wisc.edu}

\abstract{The IceCube Neutrino Observatory is uniquely sensitive to MeV neutrinos emitted during a core-collapse supernova. The Supernova Data Acquisition System (SNDAQ) monitors in real-time the detector rate deviation searching for bursts of MeV neutrinos. We present a new analysis stream that uses SNDAQ to respond to external alerts from gravitational waves detected in LIGO-Virgo-KAGRA.

\vspace{4mm}
{\bfseries Corresponding author:}
Nora Valtonen-Mattila$^{1}*$\\
{$^{1}$ \itshape 
Department of Physics and Astronomy, Uppsala University Box 516, S-75120 Uppsala, Sweden
}\\
$^*$ Speaker

\ConferenceLogo{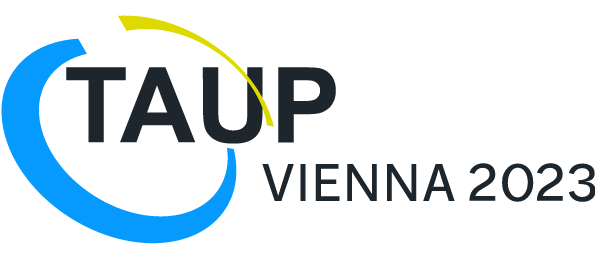}

XVIII International Conference on Topics in Astroparticle and Underground Physics 2023

\FullConference{%
XVIII International Conference on Topics in Astroparticle and Underground Physics 2023 (TAUP 2023)\\
  28 August -- 1 September, 2023\\
  Vienna, Austria}
}

\maketitle

\section{Introduction}\label{sec:Introduction}
The IceCube Neutrino Observatory~\cite{IceCube:2016zyt} is a telescope located at the South Pole, that instruments 1 km$^3$ of ice, observing neutrinos via Cherenkov radiation of energetic charged particles produced by neutrino interactions in the vicinity. The detector consists of 5160 digital optical modules (DOMs) distributed across 86 support and readout cables (strings) that are deployed deep in ice, with a spacing between them of about 125 m. While the primary purpose of IceCube is to observe high-energy (>GeV) neutrinos, it also possesses sensitivity to sufficiently bright bursts of low-energy (O(MeV)) neutrinos~\cite{IceCube:2011cwc}. 

Many transients, such as core-collapse supernovae (CCSNe) and binary systems, are expected to produce quasi-thermal neutrinos with mean energy $\expval{E_\nu(t)}\approx10$ to $30~\mathrm{MeV}$. At these energies, the neutrino interaction in ice is dominated by inverse beta decay (IBD) $\bar{\nu}_e + p \rightarrow e^+ + n$~\cite{IceCube:2011cwc}, with a few percent contribution from other interaction channels such as elastic scattering. As the positron propagates in ice, Cherenkov radiation is produced. However, due to the short average track length of the positron of $0.56~\mathrm{cm}\times(E_{e^+}/\mathrm{MeV})$, and the large spacing between strings and optical modules, a substantial portion of the photons get absorbed in ice. The resulting signal consists of single DOM hits, without directional information and competing with background noise. Each DOM has a noise rate of$~540$ Hz~\cite{IceCube:2011cwc}, with the largest contribution from radioactive decays from the glass in the pressure sphere, with a smaller contribution from atmospheric muons, which exhibit a seasonal variation~\cite{IceCube:2011cwc}. While these single hits are individually unobservable, the methodology for observing low-energy neutrinos is through the collective signal, searching for an increase in the detector rate compared to the expected background counts at a given time (see Fig.1). 

\begin{wrapfigure}{r}{0.55\textwidth}
    \includegraphics[width=\linewidth]{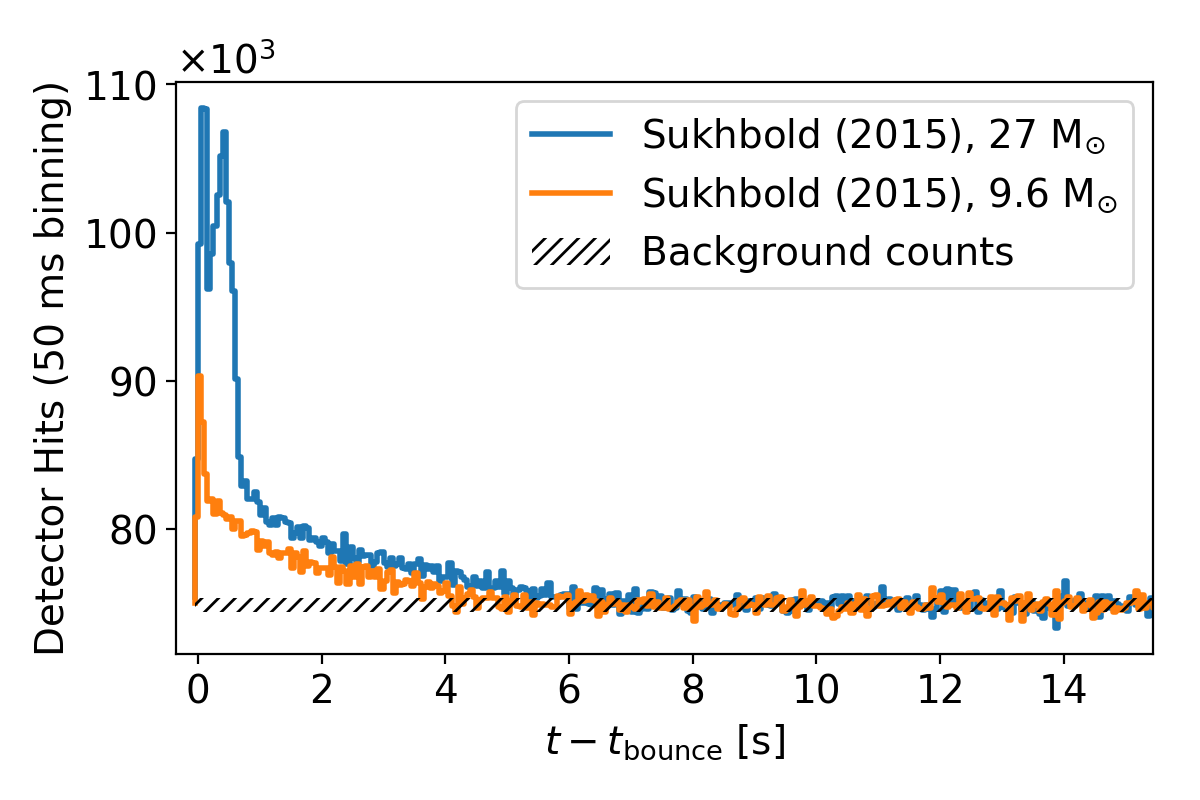}
    \caption[font=large]{Simulation of detector counts for a galactic CCSN binned in 50 ms, assuming two different masses, as a function of the time from core bounce. From Ref.~\cite{IceCube:2023MeVJakob}}
    \vspace{-9pt}
    \label{fig:rate increase}
\end{wrapfigure}

\section{Thermal Neutrinos from Transients}\label{sec:Thermal neutrinos}

Astrophysical transients produce multiple messengers which allows us to probe the properties and dynamics of the source environment. Two particularly promising sources for the detection of gravitational waves and thermal (MeV) neutrinos are CCSNe and binary systems involving neutron stars~\cite{CCSN:2020abc,Merger:2022abc}. These astrophysical phenomena are expected to be very luminous in neutrinos, with an electron anti-neutrino luminosity in the O(10$^{51}$-10$^{53}$) erg/s~\cite{Horiuchi:2018ofe, Merger:2020abc}. CCSNe, for instance, are predicted to emit a burst of neutrinos lasting $\sim$10 s, with a mean energy of $\expval{E_\nu(t)}\approx10$ to $20~\mathrm{MeV}$~\cite{Horiuchi:2018ofe}. Similarly, binary mergers involving a neutron star, such as binary neutron star (BNS) or neutron star - black hole (NSBH) systems, are expected to generate a burst of neutrinos with mean energy $\expval{E_\nu(t)}\approx10$ to $30~\mathrm{MeV}$~\cite{Merger:2020abc} over a duration of O(ms - s) (see Fig. 2).

\begin{figure}[ht]
    \centering
    \includegraphics[width=0.75\textwidth]{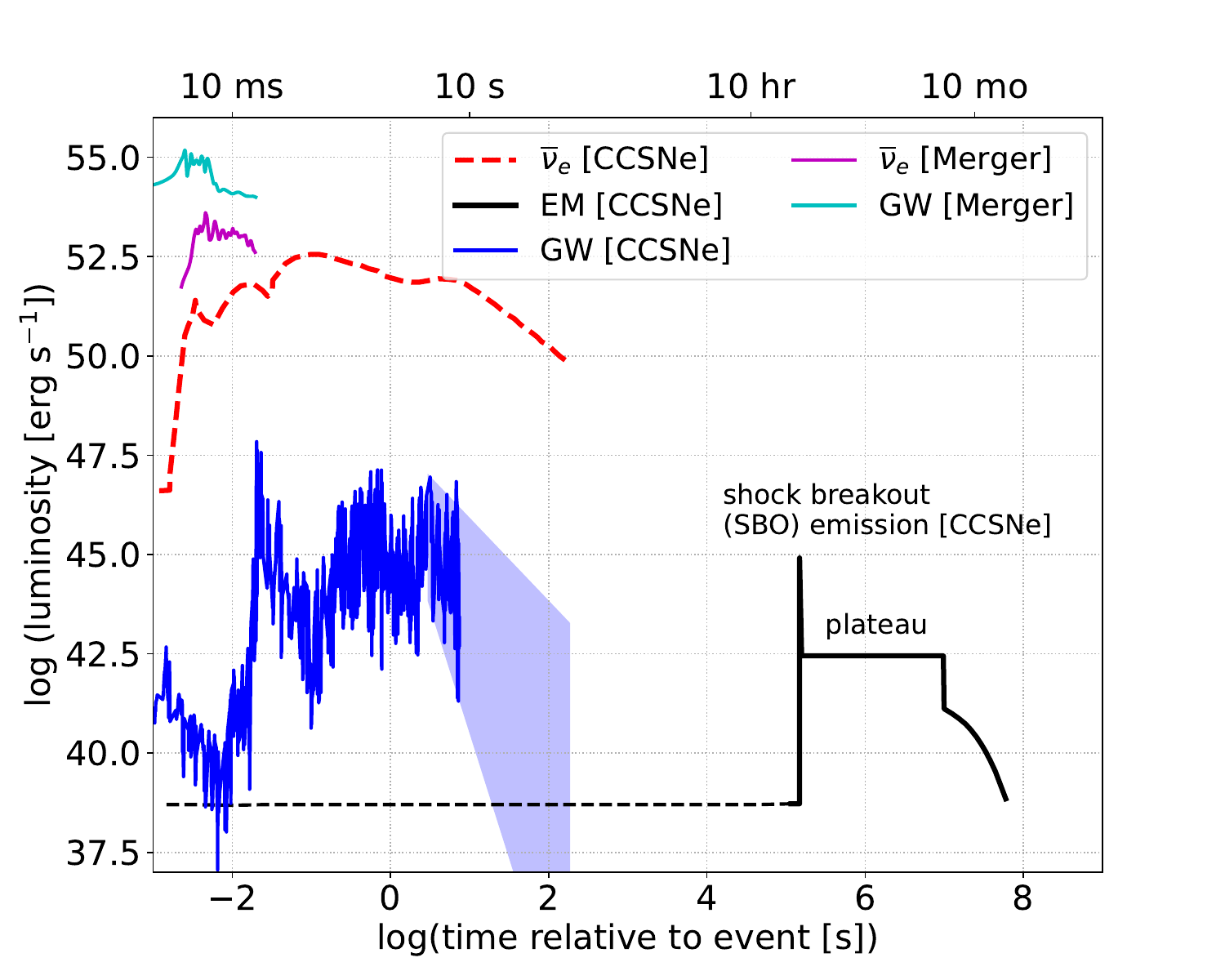}
    \caption[font=large]{Luminosity of different messengers as a function of the time relative to the to core bounce for CCSNe and relative to the merger time for binary systems. Adapted from Refs.~\cite{Merger:2016abc, Merger:2022abc}.}
    \label{fig:sndaq_time_binning}
\end{figure}


\section{Online Detection of MeV Neutrino Bursts in IceCube}\label{sec:online_system}

As previously mentioned, IceCube is sensitive to MeV neutrino bursts through the resulting rise in the single photoelectron hit rate in the DOMs. The Supernova Data Acquisition System (SNDAQ) monitors in realtime the detector rate, searching for a significant deviation $\Delta_\mu$ in all sliding search bins $r_i$ (signal window), with respect to the background mean $\mu_i$ and standard deviation $\sigma_i$ (see Fig. 3). With this, a test statistic describing the detector rate deviation can be computed, $\xi = \Delta\mu / \sigma_{\Delta\mu}$, where
\begin{align}\label{eq:delta_mu}
    \Delta\mu &= \sigma_{\Delta\mu}^2\sum_{i=1}^{N_\mathrm{DOM}}\frac{\epsilon_i\qty(r_i-\expval{r_i})}{\sigma_i^2}
    &
    &\mathrm{and}
    &
    \sigma_{\Delta\mu} &= \qty(\sum_{i=1}^{N_\mathrm{DOM}}\frac{\epsilon_i^2}{\sigma_i^2})^{-1}.
\end{align}

\noindent Here $\Delta\mu$ is the maximum-likelihood estimator of the collective rate increase across all DOMs, weighted by the relative detection efficiency $\epsilon_i$ of each DOM~\cite{IceCube:2011cwc, IceCube:2023sn}. Assuming that there is no correlated noise, this test statistic $\xi$ would follow closely a Gaussian distribution. However, due to cosmic rays, thermal noise and radioactive decay, this test statistic distribution becomes broader (see Fig. 4). In order to remove the correlated noise contributions, a deadtime of $\approx 250\, \mu$s is applied at the data acquisition stage, which reduces single DOM correlated noise, bringing the DOM rate from 540 Hz to 286 Hz. 

\begin{figure}[ht]
    \centering
    \includegraphics[width=0.95\textwidth]{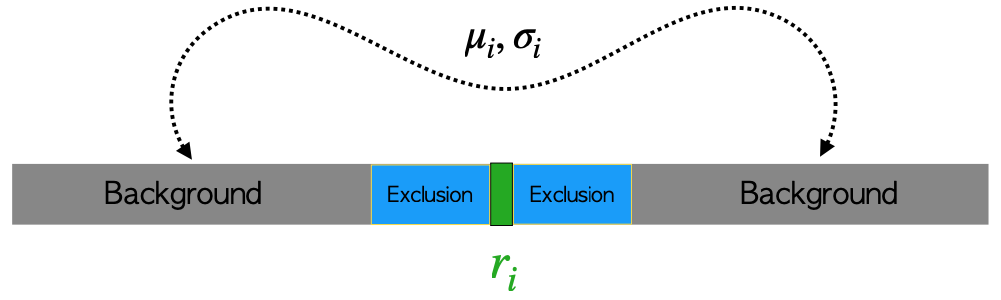}
    \caption[font=large]{Sliding window excess count search. Counts in the signal window (green) and background windows of $\pm300$~s (gray) are used to estimate $\Delta\mu$ in eq.~\eqref{eq:delta_mu}. Counts $\pm30$~s around the signal window (blue) are excluded to reduce the chance of a long signal affecting the background estimation.}
    \vspace{-10pt}
    \label{fig:sndaq_time}
\end{figure}

To further reduce the correlated noise, especially DOM-to-DOM correlations, a correction is applied to the test statistic $\xi$ at the time of the trigger. This correction aims to reduce the contribution of atmospheric muons by subtracting the contributing muon hits through a linear correction. This results in a corrected test statistic $\xi_{corr}= \xi - b \cdot R_\mu-a$, where $b$ is the slope, $a$ is the offset and $R_\mu$ is the muon hit rate. The muon hit rate is expected to vary seasonally, with a peak increase during the Austral summer, however, the contribution to the total detector rate is only a few percent~\cite{IceCube:2011cwc}.

 \begin{wrapfigure}{r}{0.53\textwidth}
    \includegraphics[width=1\linewidth]{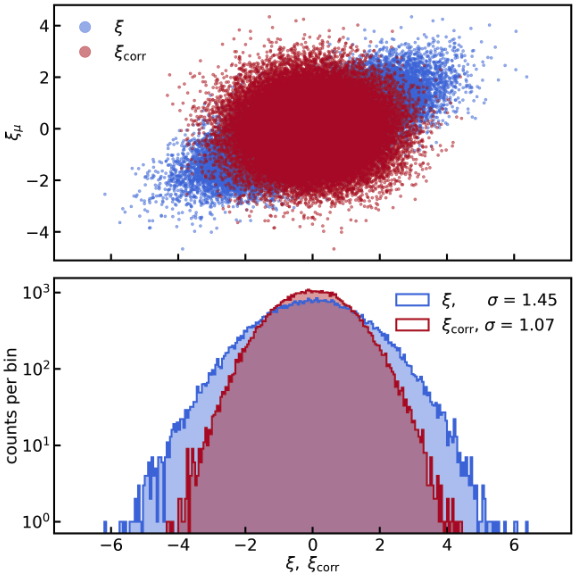}
    \caption[font=large]{Detector rate deviation test statistic $\xi$ and muon-corrected test statistic $\xi'=\xi_\text{corr}$. The top plot shows the decorrelation between the muon hit rate and the detector rate deviation when applying the correction. The bottom plot shows the histogram of the test statistics. From Ref. \cite{IceCube:2023aaa}.}
    \vspace{-10pt}
    \label{fig:rate_corr}
\end{wrapfigure}

\section{External Alert Response with SNDAQ}

\noindent Although IceCube has the possibility to perform a follow-up to external alerts using the MeV neutrino data stream~\cite{Baum:2013ekr, Baum:2015drl} it comes with a time delay of 24-72 hrs to allow for data transfer from pole to the final analysis in the northern hemisphere. The new framework utilizing SNDAQ will allow us to use external timing information to trigger analyses, allowing for a fast follow-up to external alerts, with a latency of O(10 min). This will enable to quickly follow-up on interesting alerts, such as the GRB 221009A, where an analysis of MeV neutrino data was only possible once data arrived in the north~\cite{IceCube:2023grb, IceCube:2023rhf}. The first deployment of this system will target gravitational wave alerts from the LIGO-Virgo-KAGRA O4 run, following-up alerts that are classified as bursts, since these could be potential CCSNe~\cite{LVK:2021abc}. In addition, the follow-up will also target alerts classified as Compact Binary Coalescence (CBC) with a high probability of containing a neutron star based on the alert notice parameters. 


\begin{figure}[ht]
    \includegraphics[width=\textwidth]{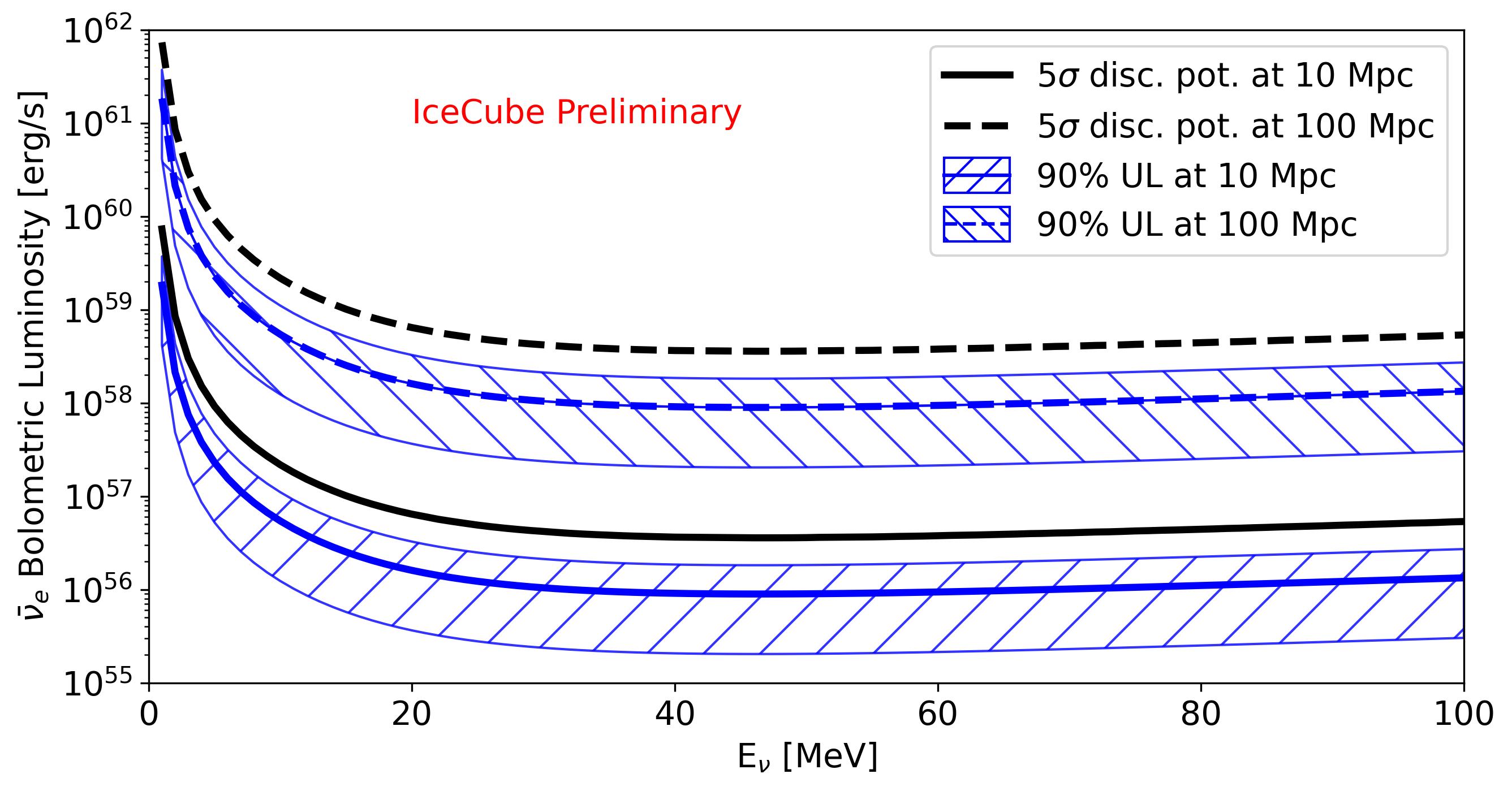}
    \caption{IceCube's MeV sensitivity to $\bar{\nu}_e$ emission from a transient at two distances; 10~Mpc (solid blue curve) and 100~Mpc (dashed blue curve). We use a 1~s search window, selecting the highest 0.5 s bin. For both distances, we plot the median 90\% upper limits expected for the bolometric luminosity (central blue curves), while the hatched regions indicate the central 68\% confidence interval on the upper limits. The $5\sigma$ discovery potential for IceCube for a source at 10~Mpc (solid black line) and 100~Mpc (dashed black line)~\cite{IceCube:2023MeV}.}
    \label{fig:exgal_sensitivity}
\end{figure}

The planned search bins are 0.5 s, 1.5 s, 4 s and 10 s, with the possibility of adding finer search bins in the future. Because the neutrino signal scales like the square inverse of the distance to the source, the flux of neutrinos must be very bright (see Fig. 5) in order for IceCube to detect transients out to distances where LVK has detected gravitational waves thus far.

\section{Conclusions}\label{sec:conclusions}

IceCube has the capability of observing bursts of MeV neutrinos by monitoring the detector rate deviation in sliding window search bins. This search is handled by SNDAQ, which monitors data in realtime with over 99\% uptime. However, the system was not designed to allow external information for triggering, effectively limiting the follow-up responses to whenever data is transmitted to the north. The new framework will enable the triggering of SNDAQ by providing externally-motivated search times, which will allow for faster external alert follow-ups with a latency of a few minutes. The first stage of deployment will target LVK O4 alerts categorized as bursts and BNS/NSBH.

\bibliographystyle{ICRC}
\bibliography{references}

\providecommand{\href}[2]{#2}\begingroup\raggedright\begin{thebibliography}{10}

\bibitem{IceCube:2016zyt}
{\bfseries IceCube} Collaboration, M.~G. Aartsen {\em et~al.} \href{http://dx.doi.org/10.1088/1748-0221/12/03/P03012}{{\em JINST} {\bfseries 12} no.~03, (2017) P03012}.

\bibitem{IceCube:2011cwc}
{\bfseries IceCube} Collaboration, R.~Abbasi {\em et~al.} \href{http://dx.doi.org/10.1051/0004-6361/201117810e}{{\em Astron. Astrophys.} {\bfseries 535} (2011) A109}.

\bibitem{IceCube:2023MeVJakob}
{\bfseries IceCube} Collaboration, J.~Beise {\em PoS} {\bfseries TAUP 2023} (2023) 159.

\bibitem{CCSN:2020abc}
E.~Abdikamalov, G.~Pagliaroli, and D.~Radice, \href{http://dx.doi.org/10.1007/978-981-15-4702-7_21-1}{{\em Gravitational Waves from Core-Collapse Supernovae}}.
\newblock Springer, Singapore, 2020.

\bibitem{Merger:2022abc}
M.~Cusinato {\em et~al.} \href{http://dx.doi.org/10.1140/epja/s10050-022-00743-5}{{\em Eur. Phys. J. A} {\bfseries 58} no.~99, (2022) }.

\bibitem{Horiuchi:2018ofe}
S.~Horiuchi and J.~P. Kneller \href{http://dx.doi.org/10.1088/1361-6471/aaa90a}{{\em J. Phys. G} {\bfseries 45} no.~4, (2018) 043002}.

\bibitem{Merger:2020abc}
Z.~Lin {\em et~al.} \href{http://dx.doi.org/10.1103/PhysRevD.101.023016}{{\em Phys. Rev. D.} {\bfseries 101} no.~023016, (2020) }.

\bibitem{Merger:2016abc}
K.~Nakamura {\em et~al.} \href{http://dx.doi.org/10.1093/mnras/stw1453}{{\em MNRAS} {\bfseries 461} no.~3, (2016) }.

\bibitem{IceCube:2023sn}
{\bfseries IceCube} Collaboration, S.~Griswold \href{http://dx.doi.org/10.48550/arXiv.2308.01843}{{\em PoS} {\bfseries ICRC2023} (2023) 1111}.

\bibitem{IceCube:2023aaa}
{\bfseries IceCube} Collaboration, R.~Abbasi {\em et~al.} \href{http://dx.doi.org/10.48550/arXiv.2308.01172}{{\em arXiv} (2023) 2308.01172}.

\bibitem{Baum:2013ekr}
{\bfseries IceCube} Collaboration, V.~Baum, D.~Heereman, and R.~Bruijn, ``{An improved data acquisition system for supernova detection with IceCube},'' in {\em {33rd International Cosmic Ray Conference}}, p.~0444.
\newblock 2013.

\bibitem{Baum:2015drl}
{\bfseries IceCube} Collaboration, V.~Baum, B.~Eberhardt, A.~Fritz, D.~Heereman, and B.~Riedel \href{http://dx.doi.org/10.22323/1.236.1096}{{\em PoS} {\bfseries ICRC2015} (2016) 1096}.

\bibitem{IceCube:2023grb}
{\bfseries IceCube} Collaboration, R.~Procter-Murphy, J.~Thwaites, B.~Brinson, N.~Valtonen-Mattila, and K.~Kruiswijk \href{http://dx.doi.org/10.48550/arXiv.2307.16354}{{\em PoS} {\bfseries ICRC2023} (2023) 1511}.

\bibitem{IceCube:2023rhf}
{\bfseries IceCube} Collaboration, R.~Abbasi {\em et~al.} \href{http://dx.doi.org/10.3847/2041-8213/acc077}{{\em Astrophys. J. Lett.} {\bfseries 946} no.~1, (2023) L26}.

\bibitem{LVK:2021abc}
{\bfseries LIGO} Collaboration, B.~Abbott {\em et~al.} \href{http://dx.doi.org/10.1103/PhysRevD.94.102001}{{\em Phys. Rev. D.} {\bfseries 94} no.~102001, (2016) }.

\bibitem{IceCube:2023MeV}
{\bfseries IceCube} Collaboration, S.~BenZvi, S.~Griswold, and N.~Valtonen-Mattila {\em PoS} {\bfseries ICRC2023} (2023) 1096.

\end{thebibliography}\endgroup

\clearpage
%
%
%


\end{document}